\documentclass[10pt]{article}
\usepackage[utf8]{inputenc}
\usepackage{mathptmx}
\usepackage{geometry}
\usepackage{authblk}
\usepackage{amssymb}
\usepackage{amsmath}
\usepackage{amsthm}
\usepackage{slashed}
\usepackage{graphicx,color,epsfig}
\usepackage{multirow}
\usepackage{cite}
\usepackage[colorlinks=true,linkcolor=red,citecolor=blue]{hyperref}
\allowdisplaybreaks[4]

\begin{document}
\renewcommand{\arraystretch}{1.2}
\newcommand{\psl}{ p \hspace{-1.8truemm}/ }
\newcommand{\nsl}{ n \hspace{-2.2truemm}/ }
\newcommand{\vsl}{ v \hspace{-2.2truemm}/ }
\newcommand{\epsl}{\epsilon \hspace{-1.8truemm}/\,  }
\def\sl{\!\!\!\!\slash}
\title{Charmless $B_s\to V S$ Decays in PQCD Approach}
\author[1]{Zhao-Wu Liu}
\author[1]{Zhi-Tian Zou$\footnote{zouzt@ytu.edu.cn}$}
\author[1,2]{Ying Li$\footnote{liying@ytu.edu.cn}$}
\author [3]{Xin Liu}
\author [4]{Jie Wang}
\affil[1]{\it \small Department of Physics, Yantai University, Yantai 264005, China}
\affil[2]{\it \small Center for High Energy Physics, Peking University, Beijing 100871, China}
\affil[3]{\it \small Department of Physics, Jiangsu Normal University, Xuzhou 221116,China}
\affil[4]{\it \small Department of Physics, Naval Aviation University, Yantai 264001,China}
\maketitle
\vspace{0.2cm}
\begin{abstract}
In this work, we investigate the $B_s\to V S$ decays  in the perturbative QCD approach, where $V$ and  $S$ denote the vector meson and  scalar meson respectively. Based on the two-quark structure, considering two different scenarios for describing the scalar mesons, we calculate the branching fractions and the direct $CP$ asymmetries of all $B_s\to VS$ decays. Most branching fractions are  predicted to be at $10^{-7}$ to $10^{-5}$, which could be measured in the LHCb and Belle-II experiments, especially for these color-allowed $B_s\to \kappa(800)(K_0^*(1430))K^*$ decays. It is found that the branching fractions of $B_s\to K_0^{*0}(1430)\bar{K}^{*0}$ and $B_s\to K_0^{*+}(1430)\bar{K}^{*-}$ are very sensitive to the scenarios, which can be used to determine whether $K_0^{*0}(1430)$ belongs to the ground state or the first excited state, if the data were available. We also note that some decays have  large direct $CP$ asymmetries, some of which are also sensitive to the scenarios, such as the $B_s \to a_0^+(1450)K^{*-}$ and the $B_s\to f_0(1500) K^{*0}$ decays. Since the experimental measurements of $B_s\to VS$ decays are on the way, combined with the available data in the future, we expect the theoretical predictions will shed light on the structure of the scalar mesons.
\end{abstract}
\section{Introduction}
In contrast to the vector and tensor mesons, the identification of the scalar mesons is still controversial, though the quark model has achieved great successes for several decades. Scalar resonances are difficult to resolve because of their large decay widths which cause a strong overlap between resonances and background, and also because several decay channels open up \cite{Close:2002zu, Amsler:2018zkm, ParticleDataGroup:2020ssz}. Furthermore, one expects non-$q\bar q$ scalar objects, like glueballs and multiquark states in the mass range below 1800 MeV. In the theoretical studies, two different scenarios \cite{Cheng:2005nb} have been often adopted to describe the scalar mesons, depending on the studies of the mass spectrum and the strong as well as eletromagnetic decays of the scalar mesons. In the so-called scenario I (S-I), all the scalar mesons are viewed as the traditional two-quark $q\bar{q}$ state, and the mesons with mass below 1 GeV, such as $a_0(980)$, $\kappa(800)$, and $f_0(500)/f_0(980)$, are the lowest lying two-quark bound states, while the mesons heavier than 1 GeV, such as $K_0^*(1430)$, $a_0(1450)$ and $f_0(1370)/f_0(1500)$, are the corresponding first excited states. In the scenario II (S-II), only those heavier mesons are interpreted as the two-quark ground states, while the light scalar mesons are the four-quark bound states or meson-meson bound states, etc. The S-II corresponds to the case that light scalar mesons are four-quark bound states, while all scalar mesons are made of two quarks in S-I. Although S-II is much preferred \cite{Cheng:2005nb, Cheng:2007st, Cheng:2013fba, Mathur:2006bs, Maiani:2004uc, tHooft:2008rus}, the final conclusion has not been confirmed yet.

Since the first $B$ meson decay mode $B\to f_0(980)K$ has been observed by Belle in 2002 through the three-body decay $K^{\pm}\pi^{\mp}\pi^{\pm}$ \cite{Abe:2002av}, which was subsequently confirmed by BaBar \cite{Aubert:2003mi},  there has been some significant progresses in the study of hadronic $B$ decays involving scalar mesons, both experimentally and theoretically. On the experimental side, there are large amounts of measurements of $B$ decays to scalar mesons reported by Belle~\cite{Garmash:2004wa, Abe:2005ig, Bondar:2004wr, Abe:2005nya}, BaBar~\cite{Aubert:2004hs, Aubert:2005wb, Aubert:2004am, Aubert:2004xg, Aubert:2004bt, Aubert:2005ce, Aubert:2004fn, Aubert:2005sk} and LHCb~\cite{LHCb:2012qnt}, which are summarized in ref.~\cite{ParticleDataGroup:2020ssz}. On the theoretical side, two-body $B$ meson decays to scalar mesons have been explored widely in different approaches, such as the QCD factorization (QCDF) \cite{Cheng:2005nb, Cheng:2007st, Cheng:2013fba, Cheng:2010sn, Li:2011kw, Li:2013aca} and the perturbative QCD approach (PQCD) \cite{Kim:2009dg, Liu:2009xm, Wang:2006ria, Shen:2006ms, Li:2019jlp, Liu:2013cvx, Liu:2019ymi, Zou:2017yxc, Zou:2017iau, Zou:2016yhb}. Combinations of the well measured experimental data with the precise theoretical predictions provide us another way to determine the nature of the scalar mesons.

It is well known to us that one cannot calculate the decay rate of two-body $B$ decay from the first principle due to the complexity of QCD. In the past decades, beyond the naive factorization approach \cite{Bauer:1986bm,Ali:1998eb, Ali:1998gb}, three major QCD-inspired approaches had been proposed to deal with the charmless non-leptonic $B$ decays, based on the effective theories, namely, QCDF \cite{Beneke:2000ry,Beneke:2003zv}, PQCD \cite{Lu:2000em, Keum:2000ph, Ali:2007ff}, and soft-collinear effective theory (SCET) \cite{Bauer:2000yr, Beneke:2002ph, Bauer:2004tj}. The difference between them is only on the treatment of dynamical degrees of freedom at different mass scales, namely the power counting. There are various energy scales involved in the hadronic $B$ decays, and the factorization theorem allows us to calculate them separately. First, the physics from the electroweak scale down to $b$ quark mass scale is described by the renormalization group running of the Wilson coefficients of effective four-quark operators. Secondly, the hard part with scale from $b$ quark mass scale to the factorization scale $\sqrt{\Lambda m_B}$ are calculated in PQCD \cite{Li:2003yj}. When doing the integration of the momentum fraction $x$ of the light quark, end point singularity will appear in the collinear factorization (QCDF and SCET) which breaks down the factorization theorem. In PQCD, the inner transverse momentum $k_T$ of the light quarks in meson is kept, which could kill the endpoint singularity. There exist also double logarithms  from the overlap of collinear and soft divergence, and the resummation of these double logarithms leads to a Sudakov form factor, which suppresses the long distance contributions and improves the applicability of PQCD. The physics below the factorization scale is non-perturbative in nature, which is described by the hadronic wave functions of mesons. They are not perturbatively calculable, but universal for all the decay processes. Since all logarithm corrections have been summed by renormalization group equations, the factorization formula does not depend on the renormalization scale. In the literatures, $B_{u,d} \to SP, SV$ and $B_{s} \to SP$ decays have been studied extensively based on PQCD, where $P$, $V$ and $S$ stand for pseudoscalar, vector and scalar mesons, respectively. For the sake of completeness, we shall study $B_s\to SV$ decays in PQCD approach in this work.


The layout of the present paper is as follows. In the next section we briefly introduce the framework of PQCD approach and wave functions of both the initial meson ($B$ meson) and the final mesons ($V$ and $S$). In Sec.\ref{sec:amplitude}, the calculation of the decay amplitude is performed in PQCD approach, and the explicit expressions are presented in this section. The numerical results and detailed discussions appear in Sec.\ref{sec:result}. Finally, in the last section we summarize this work briefly.
\section{Formalism and Wave Function}\label{sec:function}
It is convenient to study the $B$ meson decays in the $B$ meson rest framework, where two daughters of the $B$ meson fly back to back with high energy. In this case, the inner massive $b$ quark carries almost all of the energy of the $B$ meson and the massless spectator quark ($s$ quark) is soft in the initial state. In the final state, the three light quarks produced from the $b$ quark weak decay are energetic and move fast. In order to form a final meson, a hard gluon is needed to kick the soft spectator quark to make it change from soft-like to collinear-like. The hard part of the interaction becomes six-quark operator rather than four-quark. The soft dynamics here is included in the meson wave functions.

As aforementioned, there are three typical scales in $B$ meson weak decay. The first one is the $W$ boson mass $m_W$. The physics higher it is weak interaction and can be calculated perturbatively getting the Wilson coefficients $C(m_W)$ at the scale $m_W$. Using the renormalization group, we can run the Wilson coefficients from the $m_W$ scale to the ``$b$" quark mass $m_b$ scale. The physics lying between the $m_b$ and the factorizable scale $t$, namely ``hard kernel $H$" in PQCD approach, which is governed by exchanging a hard gluon, can be perturbatively calculated in PQCD approach. Finally the physics below the scale $t$ is soft and nonperturbative, which can be described by the universal hadronic wave functions of the initial and final states. According to the factorization mentioned above, the decay amplitude in PQCD approach can be written as the convolution of the Wilson coefficients $C(\mu)$, the hard kernel $H(x_i,b_i,t)$, and the initial and final hadronic wave functions:\cite{Chang:1996dw,Yeh:1997rq}:
\begin{multline}
A=\int_0^1dx_1dx_Vdx_S\int_0^{\infty}b_1db_1b_Vdb_Vb_Sdb_S\,Tr[C(t)\otimes H(x_i,b_i,t)\otimes \Phi_B(x_1,b_1)\otimes\Phi_V(x_V,b_V)\otimes\Phi_S(x_S,b_S)\\
\otimes S_t(x_i) \otimes  \exp\left[ -s(P,b) -2 \int
_{1/b}^t \frac{ d \bar\mu}{\bar \mu} \gamma_q (\alpha_s (\bar \mu))
\right], \label{eq:factorization_formula}
\end{multline}
where the $x_i(i=1,V,S)$ is the momentum fraction of the light quark in the initial and final states. The $b_i$ is the conjugate variable of the transverse momentum $k_{iT}$ of the quark in mesons. $C(t)$ are the corresponding Wilson coefficients of four quark operators, $\Phi (x)$ are the universal meson wave functions and the variable $t$ denotes the largest energy scale of hard kernel $H(x_i,b_i,t)$, which is the typical energy scale in PQCD approach and the Wilson coefficients are evolved to this scale. The exponential of $s$ function is the so-called Sudakov form factor resulting from the resummation of double logarithms occurred in the QCD loop corrections, which can suppress the contribution from the non-perturbative region. Since logarithm corrections have been summed by renormalization group equations, the  above factorization formula does not depend on the renormalization scale $\mu$ explicitly. The corrections also cause the double logarithms $\ln^2 x_i$, which can also be resummed to the jet function $S_t(x_i)$ \cite{Keum:2000wi,Lu:2000em}.

The effective Hamiltonian relevant for $B_s\to VS$  has the form \cite{Buchalla:1995vs}
\begin{eqnarray}
\label{hamiltonian}
 \mathcal{H}_{eff}=\frac{G_F}{\sqrt{2}}\Bigg\{V_{ub}^*V_{uq}(C_1(\mu)O_1(\mu)+C_{2}(\mu)O_2(\mu))-V_{tb}^*V_{tq}\sum_{i=3}^{10}C_i(\mu)O_i(\mu)\Bigg\}+{\rm h.c.},
 \end{eqnarray}
where the $V_{u(t)b}$ and $V_{u(t)q}$ are the Cabibbo-Kobayashi-Maskawa (CKM) matrix elements, with $q=d,s$ quark. The $O_{1-10}(\mu)$ are the local four-quark operators corresponding to the $B_s\to VS$ decays, which can be divided into the following three categories,
\begin{itemize}
  \item current-current (tree) operators:
\begin{eqnarray}
O_1=(\bar{b}_{\alpha}u_{\beta})_{V-A}(\bar{u}_{\beta}q_{\alpha})_{V-A},\;\;O_2=(\bar{b}_{\alpha}u_{\alpha})_{V-A}(\bar{u}_{\beta}q_{\beta})_{V-A},
\end{eqnarray}
  \item QCD penguin operators:
\begin{eqnarray}
O_3&=&(\bar{b}_{\alpha}q_{\alpha})_{V-A}\sum_{q^{\prime}}(\bar{q}^{\prime}_{\beta}q^{\prime}_{\beta})_{V-A},\;
O_4=(\bar{b}_{\alpha}q_{\beta})_{V-A}\sum_{q^{\prime}}(\bar{q}^{\prime}_{\beta}q^{\prime}_{\alpha})_{V-A},\nonumber\\
O_5&=&(\bar{b}_{\alpha}q_{\alpha})_{V-A}\sum_{q^{\prime}}(\bar{q}^{\prime}_{\beta}q^{\prime}_{\beta})_{V+A},\;
O_6=(\bar{b}_{\alpha}q_{\beta})_{V-A}\sum_{q^{\prime}}(\bar{q}^{\prime}_{\beta}q^{\prime}_{\alpha})_{V+A},
\end{eqnarray}
  \item electro-weak penguin operators:
\begin{eqnarray}
O_7&=&\frac{3}{2}(\bar{b}_{\alpha}q_{\alpha})_{V-A}\sum_{q^{\prime}}e_{q^{\prime}}(\bar{q}^{\prime}_{\beta}q^{\prime}_{\beta})_{V+A},\;
O_8=\frac{3}{2}(\bar{b}_{\alpha}q_{\beta})_{V-A}\sum_{q^{\prime}}e_{q^{\prime}}(\bar{q}^{\prime}_{\beta}q^{\prime}_{\alpha})_{V+A},\nonumber\\
O_9&=&\frac{3}{2}(\bar{b}_{\alpha}q_{\alpha})_{V-A}\sum_{q^{\prime}}e_{q^{\prime}}(\bar{q}^{\prime}_{\beta}q^{\prime}_{\beta})_{V-A},\;
O_{10}=\frac{3}{2}(\bar{b}_{\alpha}q_{\beta})_{V-A}\sum_{q^{\prime}}e_{q^{\prime}}(\bar{q}^{\prime}_{\beta}q^{\prime}_{\alpha})_{V-A},
\end{eqnarray}
\end{itemize}
where the subscripts $\alpha$ and $\beta$ are the color indices. The $q^{\prime}$ are the active quarks at the scale $m_b$, namely $q^{\prime}=u,d,s$ quarks in this work. The symbol $(\bar{b}_{\alpha}q_{\alpha})_{V-A}$ is the left handed current with the explicit expression $\bar{b}_{\alpha}\gamma_{\mu}(1-\gamma_5)q_{\alpha}$. The right handed current $(\bar{q}^{\prime}_{\beta}q^{\prime}_{\beta})_{V+A}$ has the expression $\bar{q}^{\prime}_{\beta}\gamma_{\mu}(1+\gamma_5)q^{\prime}_{\beta}$.


In PQCD, wave functions are the most important input parameters. For the initial $B_s$ meson, its wave function has been widely studied in refs.\cite{Zou:2015iwa, Ali:2007ff}. The expression can be given as
\begin{eqnarray}
\Phi_{B_s}=\frac{i}{\sqrt{2 N_c}}(\makebox[-1.5pt][l]{/}P_{B_s}+m_{B_s})\gamma_5\phi_{B_s}(x_1,b_1),
\end{eqnarray}
where the numerically suppressed term has been neglected. The $\phi_B(x_1,b_1)$ is the light-cone distribution amplitudes (LCDA) with the explicit expression as
\begin{eqnarray}
\phi_{B_s}(x,b)=N_{B}x^{2}(1-x)^{2}\exp \left[ -\frac{1}{2} \left(
\frac{xm_{B_s}}{\omega }\right) ^{2} -\frac{\omega ^{2}b^{2}}{2}\right] \label{bw} \;,
\end{eqnarray}
where the $N_B$ is the normalization constant, which can be determined by the normalization condition
\begin{eqnarray}
\int_0^1 dx \phi_{B_s}(x,b=0)=\frac{f_{B_s}}{2\sqrt{6}},
\end{eqnarray}
with the decay constant $f_{B_s}=(0.23\pm0.02)\rm{GeV}$ \cite{Zou:2015iwa,Ali:2007ff,Xiao:2006hd,Li:2004ep}. The corresponding shape parameter $\omega$ in the LCDA of the $B_s$ meson is usually taken the value $(0.5\pm0.05)\rm{GeV}$.  A model-independent determination of the $B$ meson DA from an Euclidean lattice was attempted very recently \cite{Wang:2019msf}. We also note that only the $B$ meson LCDA associated with the leading Lorentz structure in eq.~(\ref{bw}) is considered, while other power-suppressed pieces are negligible within the accuracy of the current work~\cite{Li:2012nk, Li:2014xda}.

In the two-quark picture of S-I and S-II, the two kinds of decay constants of scalar meson $S$ are defined by:
\begin{eqnarray}
\langle S(p)|\bar q_2\gamma_\mu q_1|0\rangle=f_Sp_\mu, \,\,\,\,\,\,\,\,\,\,\,
\langle S(p)|\bar q_2 q_1|0\rangle=m_S\bar {f_S}.
\end{eqnarray}
The $f_S$ and $\bar{f}_S$ are the vector and scalar decay constants respectively, which can be related with each other through the equation of motion
\begin{eqnarray} \label{eq:EOM}
\bar f_S=\mu_Sf_S, \qquad\quad{\rm with}~~\mu_S=\frac{m_S}{m_2(\mu)-m_1(\mu)},
\end{eqnarray}
$m_{1(2)}$ being the running current quark mass. It is obvious that the vector decay constant is highly suppressed by the tiny mass difference between the two running current quarks. Therefore, in contrast to the scalar one, the vector decay constant of the scalar meson, namely $f_S$, will vanish in the SU(3) limit.

As for the light scalar meson wave function, the twist-2 and twist-3 LCDAs for different components could be combined into a single matrix element \cite{Cheng:2005nb,Cheng:2007st,Cheng:2013fba}
\begin{eqnarray}
\Phi_S(x)=\frac{i}{2\sqrt{6}}[\makebox[-1.5pt][l]{/}P\phi_S(x)+m_S\phi_S^{s}(x)
+m_S(\makebox[-1.5pt][l]{/}n\makebox[-1.0pt][l]{/}v-1)\phi_S^t(x)],
\end{eqnarray}
with the twist-2 LCDA $\phi_S(x)$ and twist-3 LCDA $\phi_S^{s(t)}(x)$. The $n=(1,0,\textbf{0}_T)$ and $v=(0,1,\textbf{0}_T)$ are the light-like unit vector.
Similar to the $B_s$ meson, the LCDAs of the scalar meson also obey the normalization conditions
\begin{eqnarray}
\int_0^1 dx \phi_S(x)=f_S,\;\;\;\int_0^1 dx \phi_S^{s(t)}(x)=\bar{f}_S.
\end{eqnarray}
The twist-2 LCDA can be expanded as the Gegenbauer polynomials
\begin{eqnarray}
\phi_S(x,\mu)=\frac{3}{2\sqrt{6}}x(1-x)[f_S(\mu)+\bar{f}_S\sum_{m=1}^{\infty}B_m(\mu)C_m^{3/2}(2x-1)],
\end{eqnarray}
where $C_m^{3/2}(2x-1)$ are the Gegenbauer polynomials, and $B_m(\mu)$ are the corresponding Gegenbauer momenta. For the two twist-3 LCDAs, the asymptotic forms are adopted
\begin{eqnarray}
\phi_S^s(x)=\frac{\bar{f}_S}{2\sqrt{6}},\;\;\;\phi_S^t(x)=\frac{\bar{f}_S}{2\sqrt{6}}(1-2x).
\end{eqnarray}
The explicit values of the parameters $B_m$, $f_S$  and $\bar{f}_S$ are referred to the refs.\cite{Cheng:2005nb,Cheng:2007st,Cheng:2013fba}.

The light vector meson has the longitudinal polarization vector $\epsilon_L$ and the transverse polarization one $\epsilon_T$, whose definition can refer to refs.~\cite{Li:2005hg,Li:2004mp}. Because the another final state is a scalar meson, the transverse polarization component of the vector meson does not contribute to the decay amplitudes, required by the angular momentum conservation. The longitudinal wave function of the vector meson is given as \cite{Ball:2007rt}
\begin{eqnarray}
\Phi_V=\frac{1}{\sqrt{2N_c}}\left[m_V\makebox[-1.5pt][l]{/}\epsilon_{L}\phi_V(x)+\makebox[-1.5pt][l]{/}\epsilon_{ L}\makebox[-1.5pt][l]{/}P\phi_V^t(x)+m_V\phi_V^s(x)\right].
\end{eqnarray}
The $\phi_V(x)$ and $\phi_V^{t,s}(x)$ are the twist-2 and twist-3 LCDAs respectively, whose expressions are given as
\begin{eqnarray}
&&\phi_V(x)=\frac{3f_V}{\sqrt{6}}x(1-x)\left[1+a_{1V}^{\parallel}C_1^{3/2}(t)+a_{2V}^{\parallel}C_2^{3/2}(t)\right],\nonumber\\
&&\phi_V^t(x)=\frac{3f_V^T}{2\sqrt{6}}t^2,\;\;\phi_V^s(x)=-\frac{3f_V^T}{2\sqrt{6}}t,
\end{eqnarray}
with $t=2x-1$. The $f_V^{(T)}$ are the decay constants of the vector meson with the following values:\cite{Ball:2007rt,Ball:2006eu}
\begin{eqnarray}
&
f_{\rho}^{\parallel}=(216\pm3){\rm MeV},\;\;
f_{\omega}^{\parallel}=(187\pm5){\rm MeV},\;\;
f_{K^*}^{\parallel}=(220\pm5){\rm MeV},\;\;
f_{\phi}^{\parallel}=(215\pm5){\rm MeV},\nonumber\\
&
f_{\rho}^{\perp}=(165\pm9){\rm MeV},\;\;
f_{\omega}^{\perp}=(151\pm9){\rm MeV},\;\;
f_{K^*}^{\perp}=(185\pm10){\rm MeV},\;\;
f_{\phi}^{\perp}=(186\pm9){\rm MeV}.
\end{eqnarray}
As for the Gegenbauer moments $a_{1,2}^{\parallel}$, they have been calculated within the QCD sum rules \cite{Ball:2005vx,Misra:2007yu},  which values at $\mu=1GeV$ are given as
\begin{equation}
 a_1^{\parallel}(K^*)=0.03\pm0.02, \;\;
 a_2^{\parallel}(\rho)=0.15\pm0.07,\;\;
 a_2^{\parallel}(K^*)=0.11\pm0.09,\;\;
 a_2^{\parallel}(\phi)=0.18\pm0.08.
\end{equation}
It is noted that for these vector mesons with definite $G$ parity, ie. $\rho(\omega)$ and $\phi$, $a_1^{\parallel}=0$.

In the calculations,  the other parameters  we used are also given as \cite{ParticleDataGroup:2020ssz}.
\begin{eqnarray}
&&\Lambda_{\overline{MS}}^{f=4}=0.25\pm0.05 GeV,\;m_{B_s}=5.366 GeV,\;f_{B_s}=(0.236\pm0.03)GeV,\;\tau_{B_S}=1.527 ps,\nonumber\\
&&|V_{ud}|=0.97401\pm0.00011,\;|V_{us}|=0.22650\pm0.00048,\;|V_{ub}|=0.00361^{+0.00011}_{-0.00009},\nonumber\\
&&|V_{td}|=0.00854^{+0.00023}_{-0.00016},\;|V_{ts}|=0.03978^{+0.00082}_{-0.00060},\;|V_{tb}|=0.999172^{+0.000024}_{-0.000035},\nonumber\\
&&\alpha=(84.9^{+5.1}_{-4.5})^{\circ},\;\gamma=(72.1^{+4.1}_{-4.5})^{\circ}.
\label{parameter}
\end{eqnarray}

\section{Perturbative Calculations}\label{sec:amplitude}
\begin{figure}[!tbh]
\begin{center}
\hspace{-11cm}
\includegraphics[scale=0.2,height=7cm]{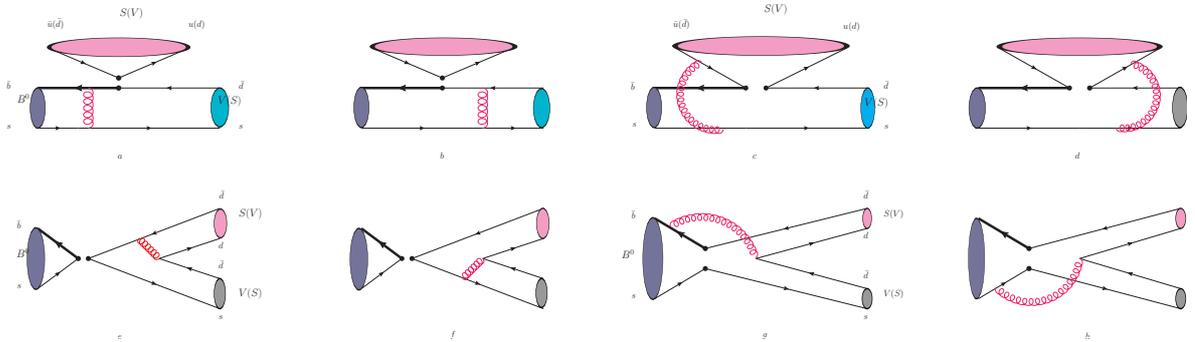}
\vspace{-2cm}
\caption{Leading order Feynman diagrams contributing to the
$B_s\,\to VS$ decays in PQCD approach}
\label{fig:diagram1}
 \end{center}
\end{figure}

In PQCD, there are eight diagrams corresponding to the $B_s\to VS$ decays at leading order, which are shown in Fig.~\ref{fig:diagram1}. The four diagrams in first line are the emission type diagrams, where the first two diagrams can be factorized into the product of the decay constant and the transition form factor, while the last two are called hard-scattering emission diagrams. The  diagrams in the second line are annihilation type diagrams. Similarly, the first two are the factorizable diagrams, while the last two ones are nonfactorizable.

Using the weak hamiltonian and the wave functions determined above,  we start to perform the perturbative calculations and present the explicit expressions of decay amplitudes corresponding to the each diagram in Fig.~\ref{fig:diagram1}. For the sake of simplicity, we define some symbols to describe the decay amplitudes with the different currents. The $LL$ denotes $(V-A)(V-A)$ current, and the $LR(SP)$ represents the $(V-A)(V+A)$ ($(S-P)(S+P)$) current.

The first two factorizable emission diagrams $(a)$ and $(b)$ in Fig.1, each decay amplitude can be separated into the decay constant of emitted meson ($M_2$) and the $B\to M_3$ transition form factor. The decay amplitudes with respect to the $LL$, $LR$ and $SP$ currents are
\begin{eqnarray}
{F}_{S}^{LL,LR}&=&8\pi C_F f_S m_B^4 \int_{0}^1dx_1dx_V\int_0^{\infty}b_1db_1b_Vdb_V\phi_B(x_1,b_1)\nonumber\\
&&\bigg\{\Big[\phi_V(x_V)(1+x_V)-r_V(2x_V-1)(\phi_V^s(x_V)+\phi_V^t(x_V))\Big]E_{ef}(t_a)h_{ef}(x_1,x_V(1-r_S^2),b_1,b_V)\nonumber\\
&&+2r_V\phi_V^s(x_V)E_{ef}(t_b)h_{ef}(x_V,x_1(1-r_S^2),b_V,b_1)\bigg\},
\\
{F}_{S}^{SP}&=&-16 \pi C_F \bar{f}_S m_B^4\int_0^1dx_1dx_V\int_0^{\infty}b_1db_1b_Vdb_V\phi_B(x_1,b_1)r_S\nonumber\\
&&\bigg\{\Big[\phi_V(x_V)+r_V\left[\phi_V^s(x_V)(x_V+2)-x_V\phi_V^t(x_V)\right]\Big]E_{ef}(t_a)h_{ef}(x_1,x_V(1-r_S^2),b_1,b_V)\nonumber\\
&&+2r_V\phi_V^s(x_V)E_{ef}(t_b)h_{ef}(x_V,x_1(1-r_S^2),b_V,b_1)\bigg\},
\end{eqnarray}
where $r_{S(V)}=m_{S(V)}/{m_{B_s}}$  and the color factor $C_F=4/3$. The explicit expressions of the functions $E_{ef}(t_{a,b})$,  $h_{ef}$ and the hard scales $t_{a,b}$ have been given in the ref.\cite{Zou:2015iwa}. The subscript ``$S$'' denotes that the scalar meson is emitted. When the vector meson is emitted, the corresponding  contributions  is also given as
\begin{eqnarray}
{F}_V^{LL,LR}&=&8\pi C_F f_Vm_B^4\int_0^1dx_1dx_S\int_0^{\infty}b_1db_1b_Sdb_S\phi_B(x_1,b_1)\nonumber\\
&&\bigg\{\Big[\phi_S(x_S)(x_S+1)+r_S(1-2x_S)[\phi_S^s(x_S)+\phi_S^t(x_S)]\Big]E_{ef}(t_a)h_{ef}(x_1,x_V(1-r_S^2),b_1,b_V)\nonumber\\
&&+2r_S\phi_S^s(x_S)E_{ef}(t_b)h_{ef}(x_V,x_1(1-r_S^2),b_V,b_1)\bigg\}.
\end{eqnarray}
We note that the $(S-P)(S+P)$ type operators can not contribute to the amplitude with a vector emitted, because the vector meson cannot be produced through this kind of operators.

For these two hard-scattering diagrams ($(c)$ and $(d)$), the decay amplitudes contain all three wave functions. The expressions corresponding to three currents are given as below
\begin{eqnarray}
{M}_S^{LL}&=&-16\sqrt{\frac{2}{3}}\pi C_F m_B^4\int_0^1dx_1dx_Sdx_V\int_0^{\infty}b_1db_1b_Sdb_S\phi_B(x_1,b_1)\phi_S(x_S)\nonumber\\
&&\bigg\{\Big[(x_S-1)\phi_V(x_V)+r_Vx_V(\phi_V^s(x_V)-\phi_V^t(x_V))\Big]E_{enf}(t_c)h_{enf}(\alpha,\beta_1,b_1,b_S)\nonumber\\
&&+\Big[(x_S+x_V)\phi_V(x_V)-r_Vx_V(\phi_V^s(x_V)+\phi_V^t(x_V))\Big]E_{enf}(t_d)h_{enf}(\alpha,\beta_2,b_1,b_S)\bigg\},\\
{M}_S^{LR}&=&16\sqrt{\frac{2}{3}}\pi C_Fr_Sm_B^4\int_0^1dx_1dx_Vdx_S\int_0^{\infty}b_1db_1b_Sdb_S\phi_B(x_1,b_1)\nonumber\\
&&\bigg\{\bigg[r_V\Big[\phi_S^s(x_S)(\phi_V^s(x_V)(x_S-x_V-1)-\phi_V^t(x_V)(x_S+x_V-1))+\phi_S^t(x_S)(\phi_V^s(x_V)(x_S+x_V-1)\nonumber\\
&&+\phi_V^t(x_V)(1+x_V-x_S))\Big]+\phi_V(x_V)(x_S-1)(\phi_S^s(x_S)+\phi_S^t(x_S))\bigg]E_{enf}(t_c)h_{enf}(\alpha,\beta_1,b_1,b_S)\nonumber\\
&&+\bigg[r_V\Big[\phi_S^s(x_S)(\phi_V^s(x_V)(x_S+x_V)+\phi_V^t(x_V)(x_V-x_S))+\phi_S^t(x_S)(\phi_V^s(x_V)(x_V-x_S)+\phi_V^t(x_V)(x_S+x_V))\Big]\nonumber\\
&&+x_S\phi_V(x_V)(\phi_S^s(x_S)-\phi_S^t(x_S))\bigg]E_{enf}(t_d)h_{enf}(\alpha,\beta_2,b_1,b_S)\bigg\},\\
{M}_S^{SP}&=&-16\sqrt{\frac{2}{3}}\pi C_F m_B^4\int_0^1dx_1dx_Vdx_S\int_0^{\infty}b_1db_1b_Sdb_S\phi_B(x_1,b_1)\phi_S(x_S)\nonumber\\
&&\bigg\{\Big[\phi_V(x_V)(x_S-x_V-1)+r_Vx_V(\phi_V^s(x_V)+\phi_V^t(x_V))\Big]  E_{enf}(t_c)h_{enf}(\alpha,\beta_1,b_1,b_S)\nonumber\\
&&+\Big[x_S\phi_V(x_V)+r_Vx_V(\phi_Vt(x_V)-\phi_V^s(x_V))\Big]  E_{enf}(t_d)h_{enf}(\alpha,\beta_2,b_1,b_S)\bigg\},
\end{eqnarray}
where the  functions $E_{enf}$, $h_{enf}$ and the inner functions can be found in ref.~\cite{Zou:2015iwa}. For the diagrams with a vector meson emitted, the decay amplitudes can be obtained by doing the following substitutes in the three decay amplitudes above
\begin{eqnarray}
\phi_{V}^{(s,t)}(x_V)\leftrightarrow \phi_S^{(s,t)}(x_S),\;x_S\leftrightarrow x_V,\,r_V\leftrightarrow r_S.
\label{substitute}
\end{eqnarray}
When the emitted meson is a vector meson, the two hard-scattering diagrams (c) and (d) will cancel with each other, because the wave function of vector meson are symmetric under the exchange $x\rightarrow(1-x)$. However, when the scalar meson is emitted, the amplitudes will not be cancelled any more, but enhanced by each other, because the LCDAs of the scalar meson are antisymmetric. So, the hard-scattering diagrams with a vector emitted are usually suppressed, and the diagrams emitting a scalar meson will provide sizable contributions or even dominate the decay amplitude, especially in those color-suppressed type decays.

For the annihilation diagrams (e) and (f), the factorizable amplitudes with a scalar meson as the $M_2$ meson flying along the $n$ direction are
\begin{eqnarray}
{A}_{S}^{LL,LR}&=&8\pi C_F f_{B_s} m_B^4\int_0^1dx_Vdx_S\int_0^{\infty}b_Vdb_Vb_Sdb_S\nonumber\\
&&\bigg\{\Big[\phi_S(x_S)\phi_V(x_V)(x_V-1)+2r_Sr_V\phi_S^s(x_S)(\phi_V^t(x_V)x_V-\phi_V^s(x_V)(x_V-2))\Big]E_{af}(t_e)h_{af}(\alpha_1,\beta,b_S,b_V)\nonumber\\
&&+\Big[x_S\phi_V(x_V)\phi_S(x_S)-2r_Vr_S\phi_V^s(x_V)(\phi_S^s(x_S)(x_S+1)
+\phi_S^t(x_S)(x_S-1))\Big]E_{af}(t_f)h_{af}(\alpha_2,\beta,b_V,b_S)\bigg\},\\
{A}_S^{SP}&=&-16\pi C_F f_{B_s} m_B^4 \int_0^1dx_Vdx_S\int_0^{\infty}b_Vdb_Vb_Sdb_S\nonumber\\
&&\bigg\{\Big[2r_S\phi_V(x_V)\phi_S^s(x_S)+\phi_S(x_S)r_V(x_V-1)(\phi_V^s(x_V)+\phi_V^t(x_V))\Big]E_{af}(t_e)h_{af}(\alpha_1,\beta,b_S,b_V)\nonumber\\
&&-\Big[\phi_V(x_V)r_Sx_S(\phi_S^t(x_S)-\phi_S^s(x_S)) +2r_V\phi_S(x_S)\phi_V^s(x_V)\Big]E_{af}(t_f)h_{af}(\alpha_2,\beta,b_V,b_S)\bigg\}.
\end{eqnarray}
Similarly, the contributions ${A}_V^{LL,LR,SP}$ from the diagrams with a vector flying along $n$ direction can be got easily from the substitutes in eq.(\ref{substitute}). The corresponding scales and hard functions are also be referred to in ref.\cite{Zou:2015iwa}.

Finally, the rest two diagrams are non-factorizable annihilation diagrams, whose amplitudes with different operators are listed as follows
\begin{eqnarray}
{W}_S^{LL}&=&-16\sqrt{\frac{2}{3}}\pi C_F m_B^4\int_0^1dx_1dx_Vdx_S\int_0^{\infty}b_1db_1b_Sdb_S\phi_B(x_1,b_1)\nonumber\\
&&\bigg\{\bigg[x_S\phi_S(x_S)\phi_V(x_V)-r_Sr_V\Big[\phi_S^s(x_S)(\phi_V^s(x_V)(x_S-x_V+3)+\phi_V^t(x_V)(1-x_S-x_V))\nonumber\\
&&+\phi_S^t(x_S)(\phi_V^s(x_V)(x_S+x_V-1)+\phi_V^t(x_V)(1-x_S+x_V))\Big]\bigg] E_{anf}(t_g)h_{anf}(\alpha,\beta_1,b_1,b_S)\nonumber\\
&&+\bigg[r_Vr_S\Big[\phi_S^s(x_S)(\phi_V^s(x_V)(x_S-x_V+1)+\phi_V^t(x_V)(x_S+x_V-1))+\phi_S^t(x_S)(\phi_V^s(x_V)(1-x_S-x_V)\nonumber\\
&&-\phi_V^t(x_V)(x_S-x_V+1))\Big]+\phi_V(x_V)\phi_S(x_S)(x_V-1)\bigg]E_{anf}(t_h)h_{anf}(\alpha,\beta_2,b_1,b_S)\bigg\},\\
{W}_S^{LR}&=&16\sqrt{\frac{2}{3}}\pi C_F m_B^4 \int_0^1dx_1dx_Sdx_V\int_0^{\infty}b_1db_1b_Sdb_S\phi_B(x_1,b_1)\nonumber\\
&&\bigg\{\Big[r_S\phi_V(x_V)(\phi_S^s(x_S)+\phi_S^t(x_S))(x_S-2)-r_V\phi_S(x_S)(\phi_V^s(x_V)-\phi_V^t(x_V))(x_V+1)\Big]
E_{anf}(t_g)h_{anf}(\alpha,\beta_1,b_1,b_S)\nonumber\\
&&+\Big[r_V\phi_S(x_S)(\phi_V^s(x_V)-\phi_V^t(x_V))(x_V-1)-r_S\phi_V(x_V)x_S(\phi_S^s(x_S)+\phi_S^t(x_S))\Big] E_{anf}(t_h)h_{anf}(\alpha,\beta_2,b_1,b_S)\bigg\},\\
{W}_S^{SP}&=&16\sqrt{\frac{2}{3}}\pi C_F m_B^4\int_0^1dx_1dx_Sdx_V\int_0^{\infty}b_1db_1b_Sdb_S\phi_B(x_1,b_1)\nonumber\\
&&\bigg\{\bigg[\phi_V(x_V)\phi_S(x_S)(x_V-1)+r_Sr_V\Big[\phi_S^s(x_S)[\phi_V^s(x_V)(x_S-x_V+3)+\phi_V^t(x_V)(x_S+x_V-1)]\nonumber\\
&&+\phi_S^t(x_S)[\phi_V^t(x_V-x_S+1)-\phi_V^s(x_V)(x_S+x_V-1)]\Big]\bigg]E_{anf}(t_g)h_{anf}(\alpha,\beta_1,b_1,b_S)\nonumber\\
&&+\bigg[x_S\phi_S(x_S)\phi_V(x_V)-r_Sr_V\Big[\phi_S^s(x_S)[\phi_V^s(x_V)(x_S-x_V+1)+\phi_V^t(x_V)(1-x_S-x_V)]\nonumber\\
&&+\phi_S^t(x_S)[\phi_V^s(x_V)(x_S+x_V-1)-\phi_V^t(x_V)(1+x_S-x_V)]\Big]\bigg] E_{anf}(t_h)h_{anf}(\alpha,\beta_2,b_1,b_S)\bigg\}.
\end{eqnarray}
Using the same substitutions as eq.~(\ref{substitute}), we get the amplitudes corresponding to the diagrams with a vector meson along the $n$ direction in final states. It is worth noting that the annihilation type diagrams can be perturbatively calculated in PQCD approach, and are the main sources of the strong phase, which are required by the direct $CP$ asymmetry.

\section{Numerical Results and Discussions}\label{sec:result}
With the above analytic decay amplitudes, we can calculate the branching fractions and direct $CP$ asymmetries of all $B_s\to VS$ decays.  Although the conventional four-quark picture \cite{Close:2002zu} for the lighter scalar meson can be used to interpret the mass degeneracy of $f_0(980)$ and $a_0(980)$,  the case that the widths of $\kappa(800)$ and $\sigma(600)$ is broader than those of $a_0(980)$ and $f_0(980)$, and the large couplings of $f_0(980)$ and $a_0(980)$ decays to $K\bar{K}$,  it is very difficult for us to predict the $B_s$ decays in the four-quark assumption due to the unknown wave functions and the decay constants beyond the conventional quark model. In Table.~\ref{br1}, we list the branching fractions and  direct $CP$ asymmetries of the $B_s\to VS$ decays involving the scalar mesons with mass below $1~ \rm{GeV}$ based on two-quark assumption. In Tables.~\ref{br2} and \ref{cp1}, the branching fractions and direct $CP$ asymmetries of the $B_s\to VS$ decays involving the heavy scalar mesons around $1.5$ GeV are presented under  S-I and S-II, respectively, though S-II is preferred \cite{Cheng:2013fba}. We note that there are not any available experimental data till now, and we thus hope that our theoretical investigations will provide some useful suggestions for experimentalists. If some decays could be measured in future, the comparison between experimental data and theoretical predictions  could shed light on the inner structures of the scalar mesons.

We acknowledge that  there  are a few of the uncertainties in the theoretical calculations, such as ones from the input parameters and from higher order radiative and power corrections. In this work,  we have considered three kinds of uncertainties. The first uncertainties are caused by the input nonperturbative parameters, namely the decay constants ($f_{B_S}$, $f_V^{(T)}$, $f_S$ and $\bar{f}_S$) of the mesons in initial and final states, the shape parameter $\omega$ in $B_s$ meson wave function and the Gegenbauer moments in the distribution amplitudes of the final mesons. One can find that this kind of errors are almost the largest one among the three kinds, therefore much precise parameters from nonperturbative approaches are called in future.  Because the $CP$ asymmetries are fractions, the uncertainties in the numerators and denominators can be cancelled by each other, leading that the $CP$ asymmetries are not sensitive to these nonperturbative parameters. The second uncertainties reflect the effects of the radiative corrections and the power corrections, which are characterized by changing the QCD scale $\Lambda_{QCD}=0.25\pm0.05$ GeV and varying the hard scale $t$ from $0.8t$ to $1.2t$. It is found that the direct $CP$ asymmetries are sensitive to these corrections, and that's because  the corrections affect the hard kernels and change the strong phases  remarkably. The third kind of uncertainties are from the CKM matrix elements shown in eq.~(\ref{parameter}).

Supposing that the light scalar mesons are the two-quark lying states, we find from Table.~\ref{br1} that  the branching fractions of  the pure annihilation decays $B_s\to a_0(980)\rho$  are  at  the order of ${\cal O}(10^{-7})$, due to the power suppression. Because $\bar{u}u$  and  $\bar{d}d$  have opposite signs in $a_0(980)$ and  $\rho^0$,  while they have same signs in $\omega$, the cancellations between contributions of $\bar u  u$ and $\bar d d $  in decay $B_s\to a_0(980)\omega$ lead its branching fraction of decay being as small as $10^{-9}$ far smaller than that of $B_s\to a_0(980)\rho^0$.  For the decay  $B_s\to a_0(980)\phi$, although the spectator strange quark can enter the $\phi$ meson, the factorizable emission diagrams have null contributions, because the scalar meson $a_0(980)$ cannot be produced through $(V\pm A)$ currents. Furthermore, the nonfactorizable emission diagrams are color-suppressed or electroweak penguin suppressed. As a result, the branching fraction of $B_s\to a_0(980)\phi$ is also at the order of ${\cal O}(10^{-7})$.

\begin{table}[!tbh]
\caption{The $CP$ averaged branching fractions (BRs) (in $10^{-6}$) and the direct $CP$ asymmetries ($\mathcal{A}_{CP}^{dir}$) (in $\%$) of the $B_s\to a_0(980)[\kappa(800)]V$ decays in PQCD approach in S-I}
 \label{br1}
\begin{center}
\begin{tabular}{l c c }
 \hline \hline
 \multicolumn{1}{c}{Decay Modes}&\multicolumn{1}{c}{BRs($10^{-6}$)} &\multicolumn{1}{c}{$\mathcal{A}_{CP}^{dir}$($\%$)}   \\
\hline\hline

 $B_s \to a_0(980)\rho^0$             &$0.55^{+0.15+0.07+0.00}_{-0.17-0.13-0.01}$  &$-18.5^{+1.8+2.0+1.1}_{-1.3-2.1-1.8}$ \\

 $B_s \to a_0(980)\omega$            &$(5.42^{+1.92+1.93+0.27}_{-1.61-1.70-0.27})\times 10^{-3}$   &$9.37^{+0.00+6.75+0.37}_{-14.2-8.96-5.08}$ \\

 $B_s \to a_0^+(980)K^{*-}$    &$2.06^{+0.41+0.45+0.05}_{-0.34-0.44-0.03}$   &$-87.9^{+7.8+5.4+2.9}_{-3.9-3.2-2.4}$  \\

 $B_s \to a_0(980)\bar{K}^{*0}$    &$4.09^{+1.58+0.67+0.17}_{-1.31-0.79-0.11}$   &$-55.1^{+5.4+8.0+0.6}_{-5.7-8.7-0.8}$   \\

 $B_s \to a_0^+(980)\rho^-$    &$0.55^{+0.14+0.14+0.01}_{-0.09-0.12-0.02}$   &$29.3^{+3.1+2.9+1.3}_{-6.7-3.8-2.2}$   \\

 $B_s \to a_0^-(980)\rho^+$    &$0.60^{+0.15+0.14+0.03}_{-0.07-0.13-0.01}$  &$-58.4^{+6.3+3.4+2.5}_{-1.7-2.1-1.4}$ \\

 $B_s \to a_0(980)\phi$         &$0.23^{+0.09+0.06+0.01}_{-0.07-0.06-0.00}$   &$-3.1^{+0.3+0.5+0.3}_{-0.5-0.5-0.1}$    \\

 $B_s \to \kappa^-(800) K^{*+}$          &$7.13^{+1.45+1.70+0.23}_{-1.38-1.19-0.14}$    &$-36.4^{+5.3+5.6+0.0}_{-5.6-7.0-0.1}$ \\

 $B_s \to \kappa^+(800)K^{*-}$        &$19.2^{+6.3+5.6+0.5}_{-5.7-3.9-0.6}$    &$6.61^{+1.22+0.30+0.50}_{-1.60-0.51-0.40}$  \\

 $B_s \to \kappa^0(800)\bar{K}^{*0}$        &$21.4^{+7.0+0.0+0.0}_{-6.3-1.7-1.1}$    & 0.0\\

 $B_s \to \bar{\kappa}^0(800)K^{*0}$    &$7.38^{+2.86+2.02+0.27}_{-2.43-1.57-0.33}$  & 0.0 \\

  $B_s \to \kappa^-(800)\rho^+$           &$34.1^{+7.4+2.4+1.6}_{-6.8-3.3-1.1}$  &$11.2^{+1.12+1.65+0.23}_{-1.40-1.51-0.34}$    \\

  $B_s \to \bar{\kappa}^0(800)\omega$        &$0.58^{+0.28+0.09+0.01}_{-0.23-0.07-0.00}$  &$-90.1^{+5.2+11.1+2.7}_{-4.9-7.5-2.2}$    \\

  $B_s \to \bar{\kappa}^0(800)\rho^0$   &$0.59^{+0.30+0.12+0.05}_{-0.23-0.09-0.03}$  &$77.8^{+5.5+8.1+1.9}_{-6.3-11.0-3.2}$   \\

  $B_s \to \bar{\kappa}^0(800)\phi$  &$0.48^{+0.20+0.15+0.08}_{-0.14-0.10-0.00}$     &0.0\\
 \hline \hline
\end{tabular}
\end{center}
\end{table}

\begin{table}[!tbh]
\caption{The $CP$ averaged branching fractions (in $10^{-6}$)  of the $B_s\to S V$ decays involving the heavier scalar mesons in PQCD approach in S-I and S-II.}
 \label{br2}
\begin{center}
\begin{tabular}{l c c }
 \hline \hline
 \multicolumn{1}{c}{Decay Modes}&\multicolumn{1}{c}{BRs(S-I)($10^{-6}$)} &\multicolumn{1}{c}{BRs (S-II)($10^{-6}$)}\\
\hline\hline

 $B_s \to a_0(1450)\rho^0$             &$5.00^{+2.32+0.80+0.28}_{-1.67-0.67-0.00}$  &$3.49^{+2.23+0.61+0.20}_{-1.51-0.45-0.06}$ \\\

 $B_s \to a_0(1450)\omega$            &$0.03^{+0.02+0.02+0.00}_{-0.01-0.01-0.00}$  &$0.02^{+0.01+0.00+0.00}_{-0.01-0.00-0.00}$  \\

 $B_s \to a_0^+(1450)K^{*-}$    &$2.25^{+0.85+0.72+0.16}_{-0.62-0.51-0.09}$   &$1.92^{+0.60+0.44+0.14}_{-0.56-0.33-0.12}$ \\

 $B_s \to a_0(1450)\bar{K}^{*0}$    &$5.94^{+2.33+1.23+0.21}_{-2.40-1.31-0.33}$ &$3.02^{+1.67+0.39+0.24}_{-1.36-0.41-0.19}$   \\

 $B_s \to a_0^+(1450)\rho^-$    &$5.21^{+2.02+0.78+0.22}_{-1.65-0.78-0.15}$ &$3.34^{+1.91+0.56+0.15}_{-1.31-0.45-0.09}$ \\

 $B_s \to a_0^-(1450)\rho^+$    &$5.19^{+1.99+0.59+0.03}_{-1.82-0.73-0.22}$ &$3.91^{+2.03+0.37+0.13}_{-1.66-0.55-0.19}$ \\

 $B_s \to a_0(1450)\phi$         &$0.42^{+0.17+0.12+0.00}_{-0.17-0.11-0.00}$&$0.14^{+0.08+0.03+0.00}_{-0.06-0.03-0.00}$  \\

 $B_s \to K_0^{*-}(1430) K^{*+}$          &$14.8^{+3.4+3.4+0.4}_{-3.6-2.6-0.6}$ &$25.7^{+8.1+4.1+0.8}_{-7.9-3.8-1.5}$ \\

 $B_s \to K_0^{*+}(1430)K^{*-}$        &$3.78^{+2.68+4.29+0.24}_{-1.85-1.98-0.29}$&$38.1^{+18.4+12.2+1.2}_{-14.5-8.6-1.1}$  \\

 $B_s \to K_0^{*0}(1430)\bar{K}^{*0}$        &$2.12^{+2.21+3.70+0.00}_{-1.24-1.06-0.13}$&$39.4^{+19.3+4.8+0.7}_{-15.1-5.2-1.9}$  \\

 $B_s \to \bar{K}^{*0}(1430)K^{*0}$    &$18.7^{+8.4+4.5+0.6}_{-6.2-2.8-0.6}$ &$30.2^{+15.3+5.1+0.6}_{-12.9-4.8-1.3}$ \\

  $B_s \to K_0^{*-}(1430)\rho^+$           &$51.8^{+12.0+3.1+2.4}_{-11.0-2.9-2.1}$ &$136^{+43+7+6}_{-39-10-4}$  \\

  $B_s \to \bar{K}^{*0}(1430)\omega$        &$1.96^{+1.07+0.41+0.16}_{-0.79-0.34-0.12}$ &$2.64^{+1.74+0.69+0.14}_{-1.42-0.52-0.15}$ \\

  $B_s \to \bar{K}^{*0}(1430)\rho^0$   &$2.35^{+1.28+0.49+0.10}_{-0.95-0.49-0.05}$ &$2.81^{+1.94+0.82+0.05}_{-1.59-0.64-0.05}$ \\

  $B_s \to \bar{K}^{*0}(1430)\phi$  &$0.76^{+0.35+0.28+0.01}_{-0.27-0.10-0.00}$  &$1.37^{+0.73+0.29+0.00}_{-0.57-0.21-0.00}$ \\

  $B_s \to f_0(1370)K^{*0} $& $3.33^{+2.16+0.77+0.13}_{-1.60-0.60-0.05}$&$1.37^{+0.90+0.27+0.04}_{-0.67-0.30-0.00}$\\

  $B_s\to f_0(1370)\rho^0$& $0.16^{+0.09+0.02+0.00}_{-0.07-0.02-0.00}$&$0.39^{+0.22+0.04+0.02}_{-0.18-0.05-0.02}$\\

  $B_s\to f_0(1370)\omega$&  $0.49^{+0.27+0.25+0.02}_{-0.21-0.19-0.03}$&$0.08^{+0.07+0.10+0.00}_{-0.03-0.04-0.00}$\\

  $B_s\to f_0(1370)\phi$& $21.0^{+13.6+8.1+0.5}_{-10.8-6.5-0.5}$&$9.09^{+5.72+1.96+0.25}_{-4.09-1.77-0.20}$\\

  $B_s\to f_0(1500) K^{*0}$ & $1.51^{+0.91+0.47+0.07}_{-0.66-0.27-0.05}$&$2.64^{+1.28+0.47+0.04}_{-1.05-0.37-0.02}$\\

  $B_s\to f_0(1500) \rho^0$& $0.29^{+0.15+0.03+0.01}_{-0.14-0.04-0.02}$&$0.87^{+0.48+0.10+0.04}_{-0.37-0.10-0.04}$\\

  $B_s\to f_0(1500) \omega$ & $ 4.58^{+2.61+0.84+0.16}_{-2.25-1.02-0.16}$&$4.91^{+4.34+0.76+0.05}_{-3.11-0.96-0.17}$\\

  $B_s\to f_0(1500) \phi$& $0.28^{+0.29+0.09+0.00}_{-0.08-0.10-0.00}$&$1.07^{+1.78+0.64+0.05}_{-0.96-0.30-0.05}$\\

  $B_s\to f_0(1710)K^{*0}$ &$0.60^{+0.37+0.15+0.02}_{-0.28-0.12-0.01}$&$0.23^{+0.14+0.00+0.00}_{-0.11-0.10-0.05}$\\

  $B_s\to f_0(1710)\rho^0$& $0.024^{+0.015+0.002+0.001}_{-0.011-0.002-0.001}$&$0.066^{+0.037+0.006+0.003}_{-0.030-0.008-0.003}$\\

  $B_s\to f_0(1710)\omega$ & $0.11^{+0.06+0.05+0.00}_{-0.04-0.04-0.00}$&$0.042^{+0.024+0.021+0.000}_{-0.016-0.014-0.000}$\\

  $B_s\to f_0(1710)\phi$ & $3.60^{+2.21+1.40+0.078}_{-1.80-1.12-0.099}$&$1.53^{+0.93+0.35+0.04}_{-0.66-0.31-0.05}$\\
 \hline \hline
\end{tabular}
\end{center}
\end{table}

\begin{table}[!t]
\caption{The direct $CP$ asymmetries ($\mathcal{A}_{CP}^{dir}$) (in $\%$) of the $B_s\to V S$ decays involving the heavier scalar mesons in PQCD approach in SI and SII}
 \label{cp1}
\begin{center}
\begin{tabular}{l c c  }
 \hline \hline
 \multicolumn{1}{c}{Decay Modes} &\multicolumn{1}{c}{$\mathcal{A}_{CP}^{dir}$(SI)($\%$)} &\multicolumn{1}{c}{$\mathcal{A}_{CP}^{dir}$(SII)($\%$)}  \\
\hline\hline

 $B_s \to a_0(1450)\rho^0$              &$-14.8^{+0.6+1.0+1.1}_{-1.1-1.7-1.3}$&$-13.7^{+1.3+2.2+1.3}_{-1.0-0.9-0.7}$ \\

 $B_s \to a_0(1450)\omega$           &$21.1^{+3.2+27+2.5}_{-5.4-6.4-0.9}$ &$30.2^{+18.6+17.2+1.4}_{-0.8-6.1-0.0}$\\

 $B_s \to a_0^+(1450)K^{*-}$    & $51.7^{+10.9+10.0+0.0}_{-15.9-6.3-1.2}$&$-93.3^{+12.2+3.9+2.0}_{-6.3-5.2-2.0}$\\

 $B_s \to a_0(1450)\bar{K}^{*0}$   &$23.8^{+7.3+6.3+0.2}_{-7.1-2.7-0.3}$&$-73.2^{+9.2+11.9+1.7}_{-9.0-11.6-1.8}$   \\

 $B_s \to a_0^+(1450)\rho^-$   &$6.62^{+1.3+0.1+0.7}_{-2.0-0.7-0.8}$&$11.02^{+3.2+0.6+0.7}_{-2.3-0.5-0.7}$   \\

 $B_s \to a_0^-(1450)\rho^+$   &$-36.1^{+2.5+2.9+1.1}_{-2.6-2.9-1.6}$&$-34.4^{+1.6+1.2+1.4}_{-1.7-1.6-1.6}$ \\

 $B_s \to a_0(1450)\phi$     &$-2.82^{+0.8+0.5+0.3}_{-0.4-0.4-0.0}$ &$-2.85^{+0.49+0.56+0.00}_{-0.71-0.86-0.00}$   \\

 $B_s \to K_0^{*-}(1430) K^{*+}$        &$8.53^{+1.82+1.13+0.00}_{-2.75-4.05-0.13}$&$0.75^{+3.47+1.78+0.00}_{-6.31-3.32-0.24}$ \\

 $B_s \to K_0^{*+}(1430)K^{*-}$       &$-8.10^{+11.56+5.29+0.84}_{-9.58-24.14-0.62}$ &$6.35^{+1.20+1.02+0.38}_{-1.54-1.19-0.48}$ \\

 $B_s \to K_0^{*0}(1430)\bar{K}^{*0}$       & 0.0&0.0\\

 $B_s \to \bar{K}^{*0}(1430)K^{*0}$    & 0.0&0.0 \\

  $B_s \to K_0^{*-}(1430)\rho^+$        &$1.37^{+0.55+2.26+0.00}_{-0.45-2.13-0.15}$&$5.24^{+0.65+0.79+0.00}_{-0.75-0.58-0.13}$    \\

  $B_s \to \bar{K}^{*0}(1430)\omega$      &$-55.5^{+3.9+18.7+1.8}_{-4.6-18.8-1.9}$ &$-85.9^{+4.4+10.0+2.6}_{-4.4-8.3-2.8}$   \\

  $B_s \to \bar{K}^{*0}(1430)\rho^0$  &$11.8^{+2.3+12.9+0.0}_{-3.5-11.9-0.7}$ &$55.0^{+12.1+9.3+0.7}_{-8.1-6.5-0.1}$  \\

  $B_s \to \bar{K}^{*0}(1430)\phi$   &0.0&0.0\\

  $B_s\to f_0(1370)K^{*0}$&  $ -4.32^{+11.17+4.47+1.74}_{-6.99-3.97-0.00}$&$10.15^{+9.39+6.75+0.26}_{-8.57-4.96-0.99}$\\

  $B_s\to f_0(1370)\rho^0$ &$-15.7^{+11.8+7.04+1.09}_{-9.9-4.33-4.06}$&$20.1^{+4.9+5.66+1.96}_{-5.0-4.59-2.19}$\\

  $B_s\to f_0(1370)\omega$ & $-31.7^{+5.4+8.1+2.87}_{-10.2-4.5-4.1}$&$-85.8^{+39.2+55.4+2.6}_{-17.6-0.7-3.0}$\\

  $B_S\to f_0(1370)\phi$& $1.22^{+0.41+0.43+0.00}_{-1.07-0.20-0.00}$&$-1.05^{+1.71+0.49+0.26}_{-1.44-0.00-0.00}$\\

  $B_s\to f_0(1500)K^{*0}$& $-78.7^{+17.0+5.8+0.8}_{-5.5-3.6-0.0}$&$68.2^{+13.7+3.7+2.0}_{-17.2-5.7-2.3}$\\

  $B_s\to f_0(1500)\rho^0$& $19.7^{10.1+9.2+2.4}_{-10.3-8.5-1.5}$&$26.6^{+5.7+5.3+2.5}_{-7.1-5.0-2.2}$\\

  $B_s\to f_0(1500)\omega$ &$11.2^{+2.8+2.7+0.2}_{-3.6-2.6-1.2}$&$15.4^{+4.8+4.9+1.0}_{-4.2-4.8-1.0}$\\

 $B_s\to f_0(1500)\phi$&$ -21.4^{+76.3+16.5+10.0}_{-32.2-31.7-0.0}$&$26.8^{+45.2+13.2+6.2}_{-11.3-5.6-0.0}$\\

 $B_s\to f_0(1710)K^{*0}$&$ -12.1^{+10.5+6.3+2.3}_{-5.9-5.3-0.0}$&$24.5^{+9.4+9.8+1.2}_{-9.0-7.5-1.9}$\\

 $B_s\to f_0(1710)\rho^0$ &$ -19.7^{+12.3+7.9+2.4}_{-9.7-2.4-0.12}$&$14.3^{+4.5+4.6+1.2}_{-4.7-3.8-1.3}$\\

 $B_s\to f_0(1710)\omega$& $ -39.1^{+8.1+1.8+3.1}_{-12.5-3.1-3.7}$&$-74.3^{+18.4+7.9+2.6}_{-7.7-2.3-2.5}$\\

 $B_s\to f_0(1710)\phi$ &$ 1.97^{+0.34+0.13+0.11}_{-0.94-0.12-0.07}$&$-2.36^{+1.93+0.58+0.36}_{-1.62-0.13-0.00}$\\
 \hline \hline
\end{tabular}
\end{center}
\end{table}

For those decays involving $\kappa(800)$ or $K^*(892)$, it is found that the branching fractions of $B_s\to K^{*0}(892)\bar{\kappa}^0(800)$ and $B_s\to K^{*+}(892)\kappa^-(800)$ decays with an emitted vector are smaller than those of $B_s\to \kappa^0(800)\bar{K}^{*0}(892)$ and $B_s\to \kappa^+(800)K^{*-}(892)$ decays with an emitted $\kappa(800)$. In fact, these four decays are controlled by transitions $ \bar b\to \bar s \bar q q $ ($q=u, d$), which is penguin dominant process. The tree operators are suppressed  by the CKM matrix elements $V_{us}V_{ub}^*$, in comparison with $V_{ts}V_{tb}^*$ corresponding to the penguin operators.  In the $B_s\to K^{*0}(892)\bar{\kappa}^0(800)$ and $B_s\to K^{*+}(892)\kappa^-(800)$ decays, the interferences between the emission type penguin contributions and the chiral enhanced annihilation contributions are constructive, while the interferences are destructive in two latter decays  $B_s\to \kappa^0(800)\bar{K}^{*0}$ and $B_s\to \kappa^+(800)K^{*-}$.

For the decays $B_s\to\rho^+\kappa^-(800)$ and $B_s\to a_0^+(980)K^{*-}$ decays that are dominated by the tree diagrams, the spectator strange quarks enter into the  $\kappa^-(800)$ and $K^{*-}$, respectively. The branching fraction of  $B_s\to\rho^+\kappa^-(800)$ is much larger than that of  $B_s\to a_0^+(980)K^{*-}$, due to  the very tiny vector decay constant of $a_0^+(980)$.  We also note that the branching fraction of $B_s \to \bar{\kappa}^0(800)\rho^0[\omega]$ is much smaller than that of decay $B_s\to a_0(980)\bar{K}^{*0}$. Because  both decays are color-suppressed processes, the emission diagrams with the $\rho^0[\omega]$ emission are suppressed by the small Wilson coefficients $C_1+C_2/3$. Furthermore, for $B_s\to\bar{\kappa}^0(800)\rho^0[\omega]$ decay, the contributions from nonfactorizable emission diagrams ($c$) and ($d$) are cancelled by each other, while in decay $B_s\to a_0^+(980)K^{*-}$ the two diagrams strengthen with each other and are enhanced by the large Wilson coefficient $C_2$. These two reasons lead to this large difference.  At last, the pure penguin type $B_s\to \bar{\kappa}^0(800)\phi(1020)$ decay has a very small branching fraction, because it is suppressed by the small CKM elements $V_{td}V^*_{tb}$.

In Table.~\ref{br1}, the direct $CP$ asymmetries of concerned decay modes are also presented. One finds that there are three large $CP$ asymmetries in  $B_s\to a_0^+(980)K^{*-}$, $B_s\to \bar{\kappa}^0\rho^0$ and $B_s\to \bar{\kappa}^0 \omega$. For the $B_s\to a_0^+(980)K^{*-}$ decays, the contributions from tree operators in emission diagrams are suppressed by the tiny vector constant of $a_0^+(980)$, which leads to that the penguin contributions are comparable with the tree contributions. Thus, the large $CP$ asymmetry in this decay is sizable, because the direct $CP$ asymmetry is proportional to the interference between the tree and penguin contributions. Similarly,   the tree contributions of $B_s\to \bar{\kappa}^0\rho^0[\omega]$ are suppressed by the small Wilson coefficients $C_1+C_2/3$, and the large $CP$ asymmetries in these two decays are understandable.  However,  for the decay $B_s\to a_0(980)\bar{K}^{*0}$, the nonfactorizable emission diagrams with the scalar meson $a_0(980)$ emission are enhanced by the large Wilson coefficient $C_2$, so the contributions from tree operators dominate the decay amplitude.  So,  the $CP$ asymmetry of $B_s\to a_0(980)\bar{K}^{*0}$ is smaller than the corresponding decay $B_s\to \bar{\kappa}^0\rho^0$. In short, the fact that the large emission diagrams with a scalar meson emitted are suppressed by the tiny vector decay constant of scalar meson or forbidden required by the charge conjugation invariance or conservation of vector current leads  to the large direct $CP$ asymmetries.  The similar phenomena also appear in the corresponding $B_{u,d}\to VS$ decays. For instance, the direct $CP$ asymmetries of $B^{-(0)}\to f_0(980)\rho^{-(0)}$ decays in ref.~\cite{Cheng:2013fba} are almost $70\%$. In addition, for the penguin operator dominated decays, such as $B_s\to \kappa^+K^{*-}$ and $B_s \to \bar{\kappa}^0(800)\phi$,  the $CP$ asymmetries are about zero, as we expected. In the experimental side, some decays with both large branching fractions and large direct $CP$ asymmetries, such as the $B_s\to a_0^{+}(980)[a_0^{+}(1450)]K^{*-}$ decays, are suggested to be measured in the experiments.

For $f_0(980)$ and $\sigma$, the experimental data of $D_s^+\to f_0(980)\pi$, $\phi\to f_0(980)\gamma$ and the observed relation $\Gamma(J/\psi\to f_0(980)\omega)\simeq\frac{1}{2}\Gamma(J/\psi\to f_0(980)\phi)$ imply that the $f_0(980)$ and $\sigma$ have a similar mixing  as the $\eta$-$\eta^{\prime}$, so under two-quark assumption we define
\begin{equation}
\left(
\begin{array}{c}
 \sigma\\
 f_0(980)
\end{array}
\right )
=
\left(
\begin{array}{cc}
 \cos\theta & -\sin\theta \\
 \sin \theta & \cos\theta
\end{array}
\right )
\left(
\begin{array}{c}
 \bar{n}n\\
 \bar{s}s
\end{array}
\right ),
\end{equation}\label{angle}
with $\bar{n}n=(\bar{u}u+\bar{d}d)/\sqrt{2}$. The mixing angle $\theta$ has not been well determined by current experimental measurements, though there are various constraints from different measurements. In  refs.\cite{Cheng:2002ai, Anisovich:2002wy, Gokalp:2004ny}, the authors summarized the current experiments data, such as the branching fractions of $J/\psi\to f_0\omega$, $J/\psi\to f_0\phi$ \cite{Anikina:2000tj} and the couplings of $f_0\to \pi\pi/ K K$ \cite{KLOE:2002deh,Achasov:2000ym, CMD-2:1999znb, WA102:1999fqy}, and constrained the mixing angle $\theta$  to be in the ranges of $[25^\circ,40^\circ]$ and $[140^\circ, 165^{\circ}]$. A LHCb measurement of the upper limit on the branching fraction product ${B}({B}^0\to J/\psi f_0(980))\times{B}(f_0(980)\to \pi^+\pi^-)$ constrains the mixing angle $\mid\theta\mid<30^{\circ}$\cite{Czakon:2014xsa}. In ref.~\cite{Cheng:2013fba},  $B^-\to f_0(980)K^-/f_0(980)K^{*-}$ can be accommodated with $\theta$ in the vicinity of $20^{\circ}$, and the mixing angle is adopted as $17^{\circ}$. In short, we can not obtain a universal mixing angle $\theta$ to accommodate to all the current experimental measurements, and we therefore set the mixing angle $\theta$ to be a free parameter in current work.

\begin{figure}[!tbh]
\begin{center}
\includegraphics[scale=0.8]{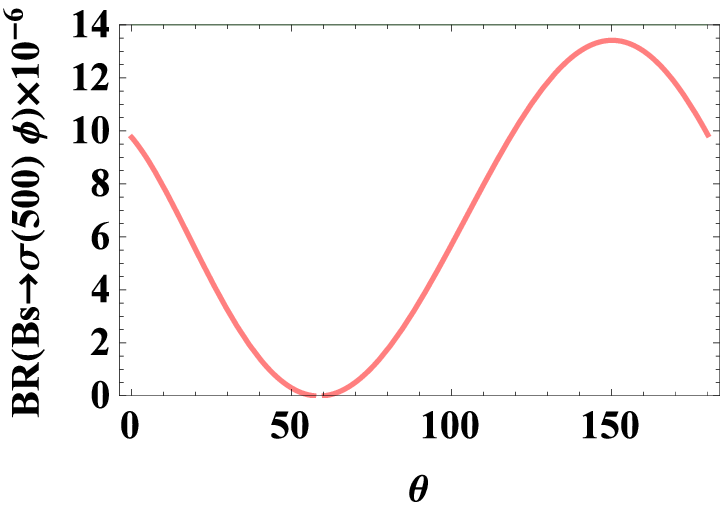}\,\,\,\,\,\,\,\,
\includegraphics[scale=0.8]{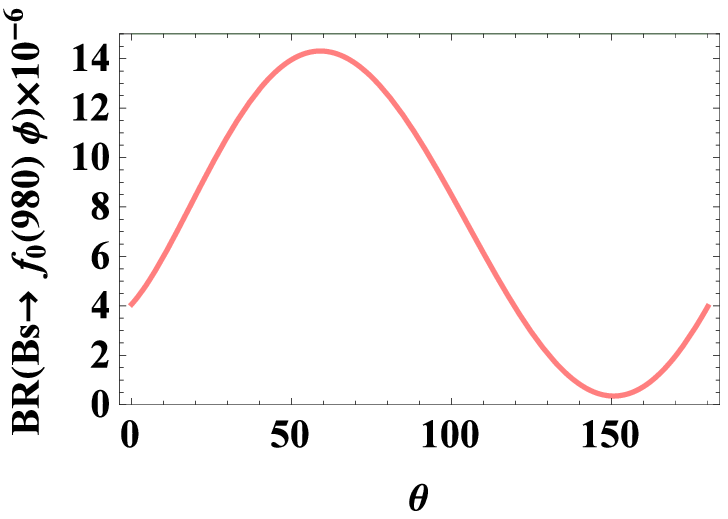}
\includegraphics[scale=0.8]{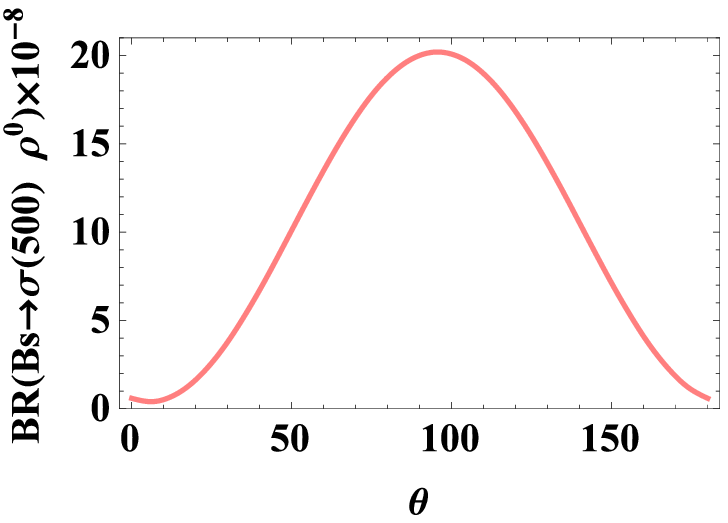}\,\,\,\,\,\,\,\,
\includegraphics[scale=0.8]{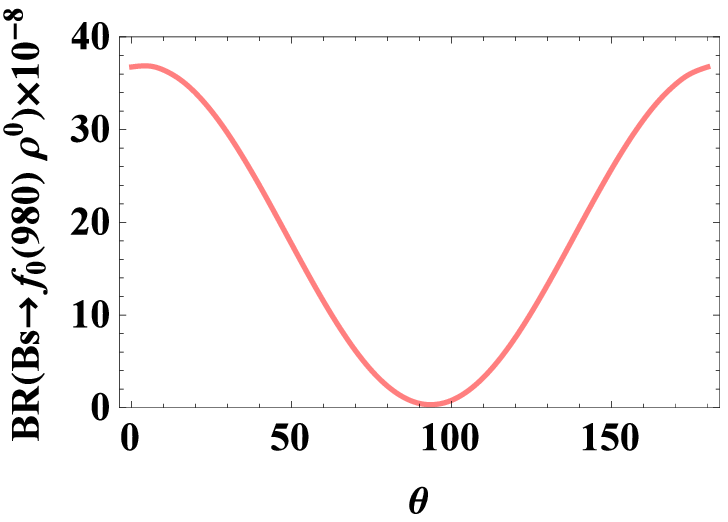}
\includegraphics[scale=0.8]{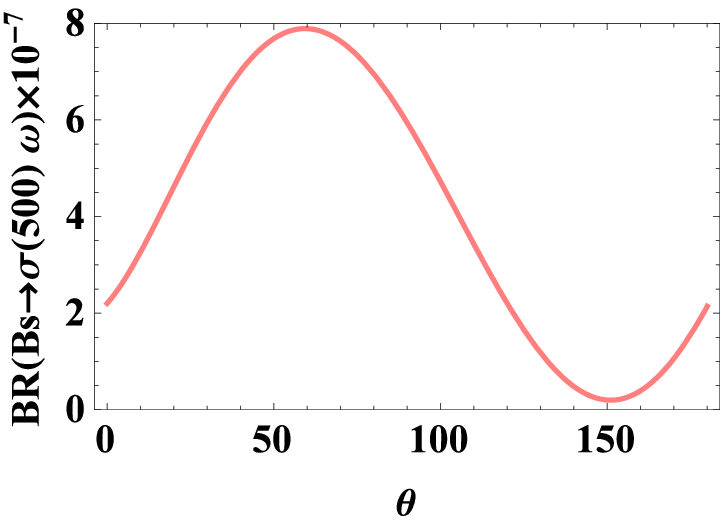}\,\,\,\,\,\,\,\,
\includegraphics[scale=0.8]{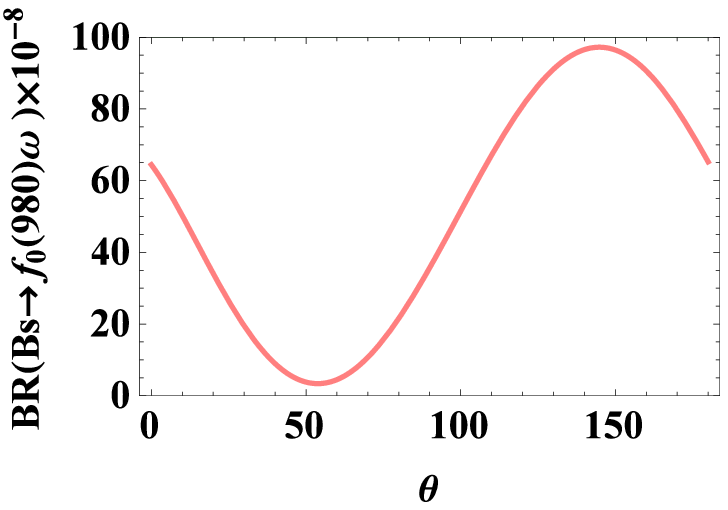}
\includegraphics[scale=0.8]{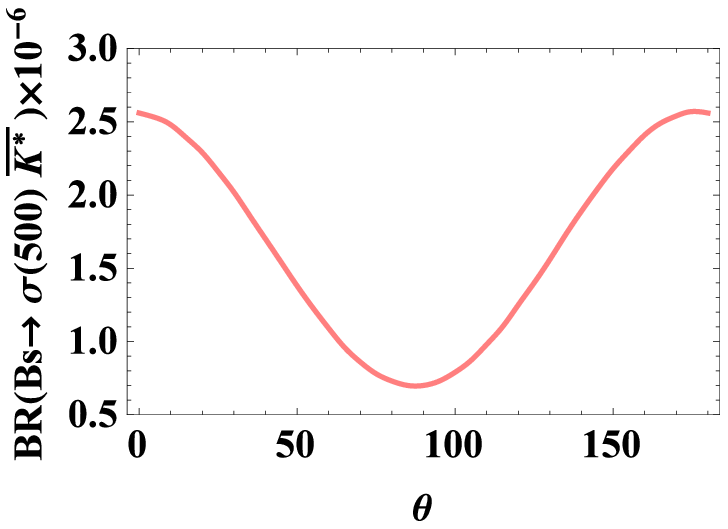}\,\,\,\,\,\,\,\,
\includegraphics[scale=0.8]{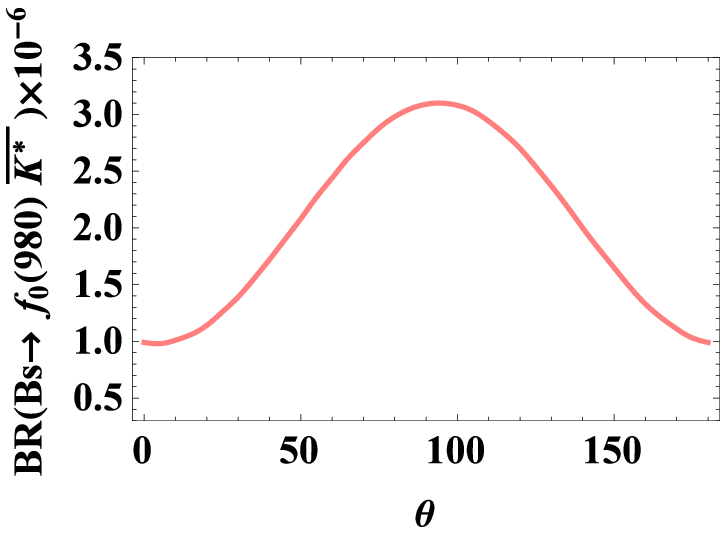}
\caption{The branching fractions of the $B_s\to f_0(980)[\sigma]V$ decays versus the $f_0(980)-\sigma$ mixing angle $\theta$.}
\label{fig:diagram2}
 \end{center}
\end{figure}

\begin{figure}[!tbh]
\begin{center}
\includegraphics[scale=0.8]{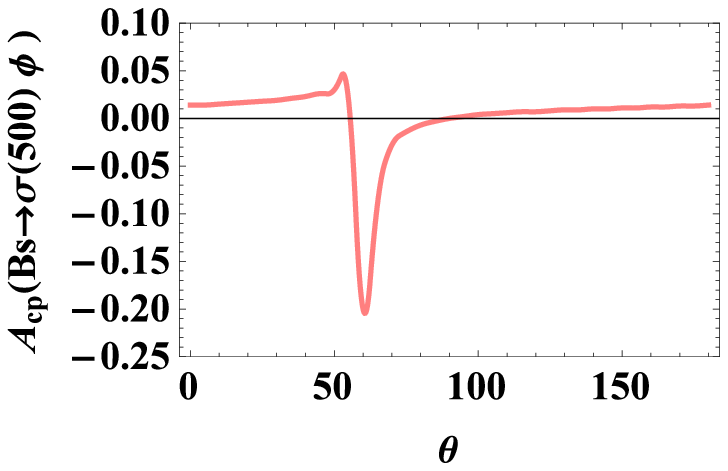}\,\,\,\,\,\,\,\,
\includegraphics[scale=0.8]{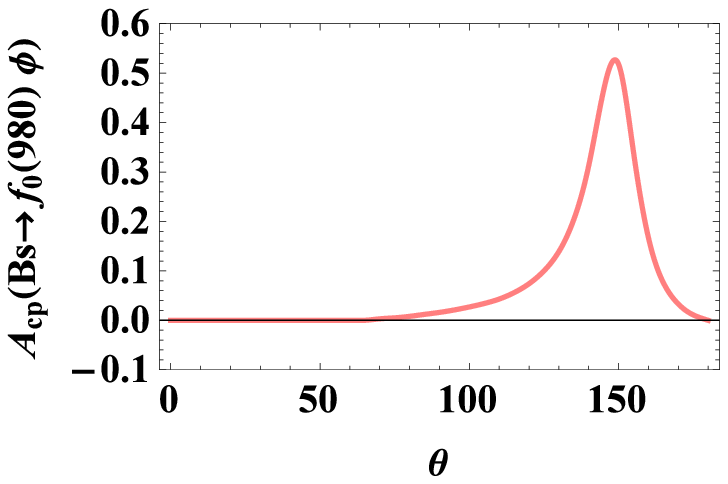}
\includegraphics[scale=0.8]{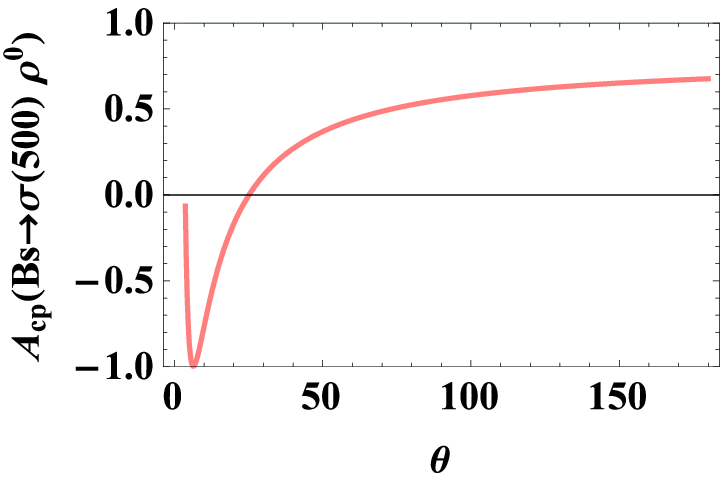}\,\,\,\,\,\,\,\,
\includegraphics[scale=0.8]{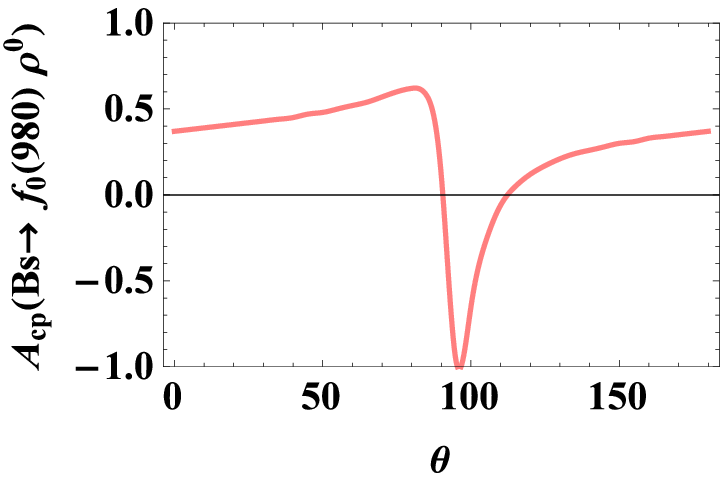}
\includegraphics[scale=0.8]{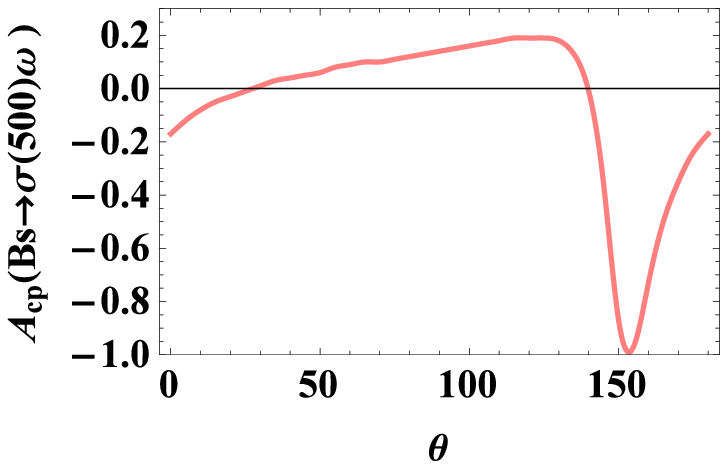}\,\,\,\,\,\,\,\,
\includegraphics[scale=0.8]{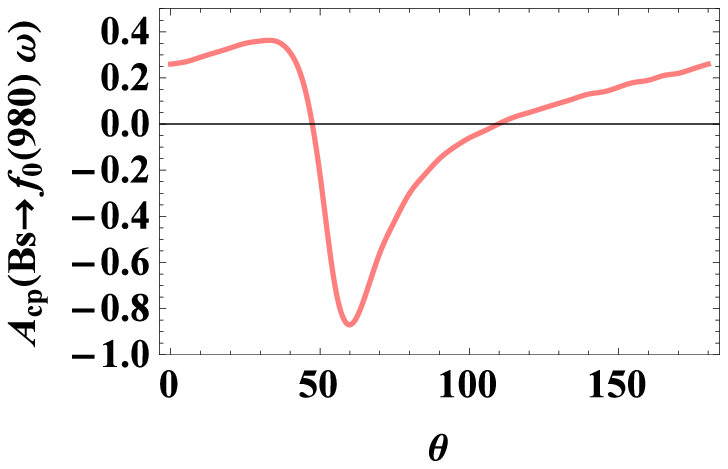}
\includegraphics[scale=0.8]{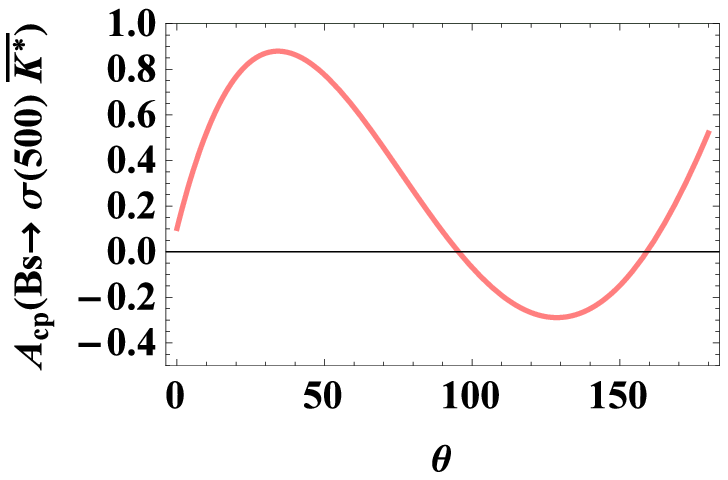}\,\,\,\,\,\,\,\,
\includegraphics[scale=0.8]{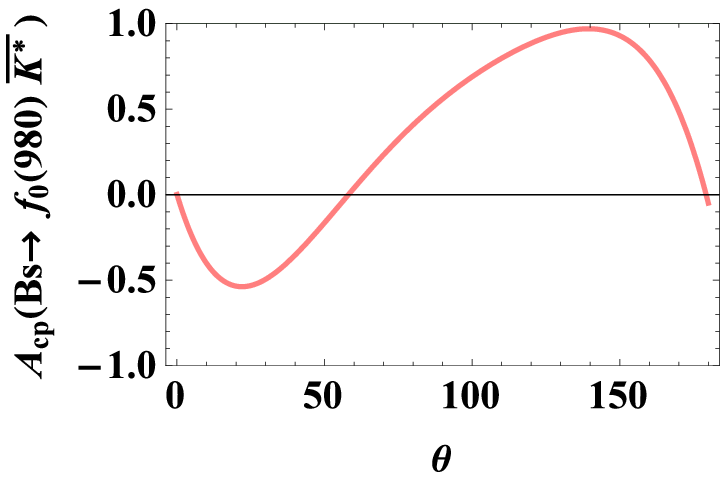}
\caption{The direct $CP$ asymmetries of the $B_s\to f_0(980)[\sigma]V$ decays versus the $f_0(980)-\sigma$ mixing angle $\theta$.}
\label{fig:diagram3}
 \end{center}
\end{figure}
In the Figure.~\ref{fig:diagram2} and Figure.~\ref{fig:diagram3}, we plot the branching fractions and the direct $CP$ asymmetries of the $B_s\to f_0(980)[\sigma]V$ decays versus the $f_0(980)-\sigma$ mixing angle $\theta$ defined in eq.(\ref{angle}), respectively. When the experimental data are available, these curves could be used to determine the mixing angle $\theta$. For example, if an obvious direct $CP$ asymmetry of $B_s\to f_0(980)\phi $ decay were measured, the $\theta$ would be an obtuse angle. In fact, for the decay $B_s\to f_0(980)\phi$, when the mixing angle is about $145^{\circ}$, the total penguin contribution is highly suppressed by the cancellation between the two type of penguin contributions from $\bar{s}s$ component and $\bar{n}n$ component, respectively, thus it is comparable to the small tree contribution from the $\bar{n}n$ component. The remarkable interference between those two contributions leads to a large $CP$ asymmetry. Similarly, a large $CP$ asymmetry in $B_s\to \sigma\phi $ denotes that the mixing angle $\theta$ is an acute angle. The reason is very similar to that of $B_s\to f_0(980)\phi $ decay.

In refs.\cite{Zou:2020ool,Li:2019jlp}, we have studied the two-body $B\to K^{*}_{0,2}(1430)f_0(980)/\sigma$ decays and the three-body $B_{(s)}\to \phi\pi\pi$ decays with the $f_0(980)$ as the resonance in detail. Based on PQCD predictions and the measurements from the BaBar \cite{BaBar:2011ryf} and LHCb \cite{LHCb:2016vqn}, if $\theta \sim 145^{\circ}$,  the theoretical predictions of the branching ratios are in good agreement with  experimental measurements, which indicates that an obtuse angle is preferred. Here, since there are no any available data, we also adopt the $\theta=145^{\circ}$ and list the predictions as
\begin{eqnarray}
{B}(B_s\to\sigma \bar{K}^{*0})&=&(2.0^{+0.8+0.3+0.1}_{-0.7-0.4-0.0})\times 10^{-6},\nonumber\\
{B}(B_s\to\sigma \rho^0)&=&(8.8^{+3.9+1.6+0.4}_{-3.6-1.6-0.3})\times10^{-8},\nonumber\\
{B}(B_s\to \sigma\omega)&=&(2.8^{+2.4+1.3+0.4}_{-0.7-0.3-0.3})\times10^{-8},\nonumber\\
{B}(B_s\to \sigma \phi)&=&(13.3^{+5.7+3.6+0.3}_{-4.4-3.3-0.4})\times10^{-6},\nonumber\\
{B}(B_s\to f_0(980)\bar{K}^{*0})&=&(1.8^{+0.6+0.3+0.1}_{-0.6-0.3-0.1})\times10^{-6},\nonumber\\
{B}(B_s\to f_0(980)\rho^0)&=&(2.3^{+1.1+0.3+0.1}_{-0.8-0.2-0.1})\times10^{-7},\nonumber\\
{B}(B_s\to f_0(980)\omega)&=&(9.7^{+5.0+1.5+0.0}_{-4.2-1.4-0.3})\times10^{-7},\nonumber\\
{B}(B_s\to f_0(980)\phi)&=&(4.8^{+3.4+2.4+0.2}_{-2.2-1.5-0.0})\times10^{-7}.
\end{eqnarray}

Although recent study in ref.\cite{Cheng:2013fba} indicated that the S-II is favored for describing the heavier scalar mesons, we still calculated the branching fractions and the direct $CP$ asymmetries of the $B_s\to VS$ decays involving a heavier scalar meson with the mass around $1.5$ GeV under both scenarios, and the numerical results are  summarized in the Tables.~\ref{br2} and \ref{cp1}, respectively. From the tables, it is found that the branching fractions of decays with heavy scalar are a little larger than those with light one. Because under two-quark assumption, the light scalar mesons and the corresponding heavy ones have same quark components, the situation of the decays with heavier scalar mesons are similar to the decays with light ones. For the sake of simplicity, we will not discuss them any more. As the heavier scalar mesons can be described in two possible scenarios, we would like to check which scenario is preferable based on our predictions. One can find from Table.~\ref{br2} that the branching fractions of  $B_s \to K_0^{*}(1430)K^{*}$ under S-II are much larger than those under S-I. For example, the branching fraction of charged channel $B_s \to K_0^{*+}(1430)K^{*-}$ under S-II is about $3.8\times 10^{-5}$, which is larger than the result based on S-I by almost one order of magnitude. Therefore, the future measurements of these four decays in LHCb or Belle-II will help us to differentiate two scenarios. Besides the branching fractions, we also find that  the direct $CP$ asymmetries of some decays under the different scenarios vary greatly, for example, the  large $CP$ asymmetries of $B_s\to a_0^+(1450)K^{*-}$ and $B_s\to f_0(1500)K^{*0}$ decays  have different sign under the two scenarios. If the large amount of experimental data continues to accumulate, the comparison between our predictions and future measurements could also help us to study the nature of scalar particles.

\section{Summary}
In this work we studied $B_s \to VS$ decays within the framework of PQCD approach, where $S$ denotes a scalar meson. Under two scenarios for describing the nature of the scalar meson, we calculated the branching fractions and the direct $CP$ asymmetries of these decay modes. It is shown that the branching fractions of some decays are at the order of $10^{-6}$, which can be measured  in the running  experiments such as the LHCb and the Belle-II. We also note that some decays have large direct $CP$ asymmetries, such as the $B_s\to a_0^+K^{*-}$ decays,  because the contributions with a scalar meson emission are suppressed or vanish. For the decays involving heavier scalar meson, the decays $B_s \to K_0^{*}(1430)K^{*}$ could be used to distinguish which scenario is preferable for describing the scalar mesons.

\section*{Acknowledgment}
This work is supported in part by the National Science Foundation of China under the Grant Nos. 11975195, 11705159, 11765012 and 11205072 and the Natural Science Foundation of Shandong province under the Grant No.ZR2019JQ04 and ZR2018JL01 This work is also supported by the Project of Shandong Province Higher Educational Science and Technology Program under Grants No. 2019KJJ007. X.L. is supported in party the Research Fund of Jiangsu Normal University under Grant No.~HB2016004.
\bibliographystyle{bibstyle}
\bibliography{mybibfile}
\end{document}